\documentclass[conference, 9pt]{IEEEtran}
\IEEEoverridecommandlockouts
\usepackage{cite}
\usepackage{threeparttable}
\usepackage{microtype}
\usepackage{graphicx}
\usepackage{subfigure}
\usepackage{booktabs} 
\usepackage{algorithm}
\usepackage{algorithmic}
\usepackage{hyperref}
\usepackage{amsmath,amssymb,amsfonts}
\usepackage{textcomp}
\usepackage{xcolor}
\usepackage{makecell}
\usepackage{tabularx}
\usepackage{fancyhdr}
\usepackage{marvosym}
\usepackage{multirow}

\def\BibTeX{{\rm B\kern-.05em{\sc i\kern-.025em b}\kern-.08em
    T\kern-.1667em\lower.7ex\hbox{E}\kern-.125emX}}
\begin{document}

\title{From Quarter to All: Accelerating Speculative LLM Decoding via Floating-Point Exponent Remapping and Parameter Sharing}

\author{\IEEEauthorblockN{Yushu Zhao$^{*}$, Yubin Qin$^{*}$, Yang Wang\textsuperscript{\Letter}, Xiaolong Yang, Huiming Han, Shaojun Wei, Yang Hu, Shouyi Yin\textsuperscript{\Letter}}
\IEEEauthorblockA{BNRist, Tsinghua University}
\IEEEauthorblockA{Email: wangyang\_imec@mail.tsinghua.edu.cn, yinsy@tsinghua.edu.cn}
\thanks{* Equal contribution.}
}

\maketitle

\begin{abstract}
Large language models (LLMs) achieve impressive performance across diverse tasks but exhibit high inference latency due to their large parameter sizes. While quantization reduces model size, it often leads to performance degradation compared to the full model. Speculative decoding remains lossless but typically incurs extra overheads. We propose SPEQ, an algorithm–hardware co-designed speculative decoding method that uses part of the full-model weight bits to form a quantized draft model, thereby eliminating additional training or storage overhead. A reconfigurable processing element array enables efficient execution of both the draft and verification passes. Experimental results across 15 LLMs and tasks demonstrate that SPEQ achieves speedups of 2.07×, 1.53×, and 1.45× compared over FP16, Olive, and Tender, respectively.

\end{abstract}
\begin{IEEEkeywords}
large language model, speculative decoding, model quantization, algorithm-hardware co-design.
\end{IEEEkeywords}

\section{Introduction}
\begin{figure*}[t]
    \centering
    \includegraphics[width=0.95\textwidth,keepaspectratio]{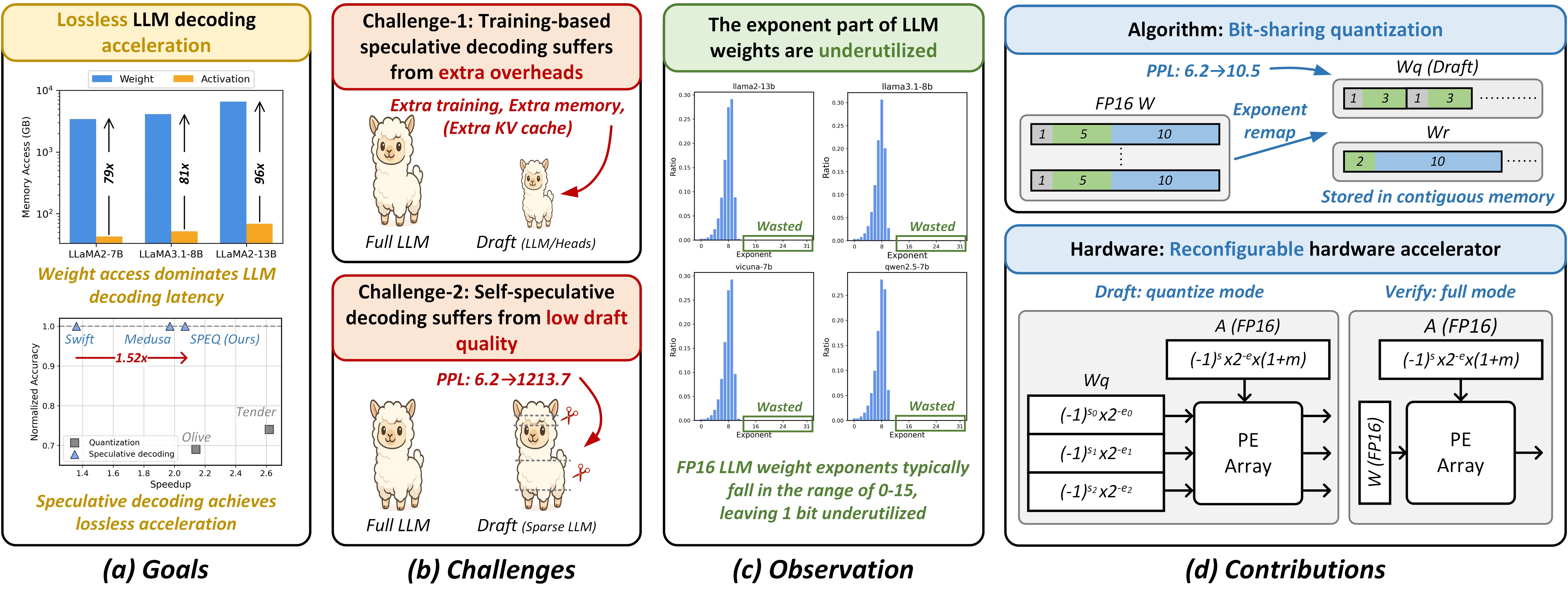} 
    \caption{Overview of SPEQ, an algorithm-hardware co-design approach for accelerating LLM speculative decoding.}
    \label{fig_overview}
\end{figure*}

Large language models (LLMs) have demonstrated exceptional performance across a wide range of downstream tasks and real-world applications, including knowledge tasks, reasoning tasks, and conversational tasks \cite{DBLP:journals/corr/abs-2303-08774}. 
The escalating scale of LLMs, which now often exceed hundreds of billions of parameters \cite{DBLP:journals/corr/abs-2307-09288,DBLP:journals/corr/abs-2407-21783}, 
introduces significant weight-loading latency during inference, as shown in Fig. \ref{fig_overview}(a). This latency acts as a primary bottleneck, degrading overall inference speed and posing a critical obstacle for real-time applications that require rapid responses.

Model compression techniques have been proposed to address the substantial memory requirements of LLMs, with quantization standing out as one of the most effective methods. By reducing the numerical precision of model weights from FP16 to lower bit-width formats, quantization significantly shrinks model size. Propelled by recent algorithmic innovations \cite{DBLP:conf/mlsys/0002TTYCWXDG024, DBLP:journals/corr/abs-2210-17323, DBLP:conf/isca/0003THL00LG023, DBLP:conf/isca/LeeLS24,DBLP:conf/icml/XiaoLSWDH23}, the performance of quantized LLMs on GPUs and specialized hardware accelerators has seen marked improvement.
Nevertheless, two critical challenges limit the widespread adoption of quantization. \textbf{ (1) Quantization introduces cumulative errors in long-sequence generation}, degrading performance compared to the full-precision counterpart. As a case in point, 4-bit AWQ \cite{DBLP:conf/mlsys/0002TTYCWXDG024} incurs a significant 5.4\% accuracy loss on the MATH task \cite{DBLP:conf/nips/HendrycksBKABTS21, DBLP:journals/corr/abs-2501-03035}. \textbf{(2) Quantization modifies the model's output distribution}, creating a potential misalignment that can compromise its effectiveness and safety in real-world applications \cite{DBLP:conf/nips/EgashiraVSHV24}.

\begin{figure}[t]
    \centering
    \includegraphics[width=\columnwidth,keepaspectratio]{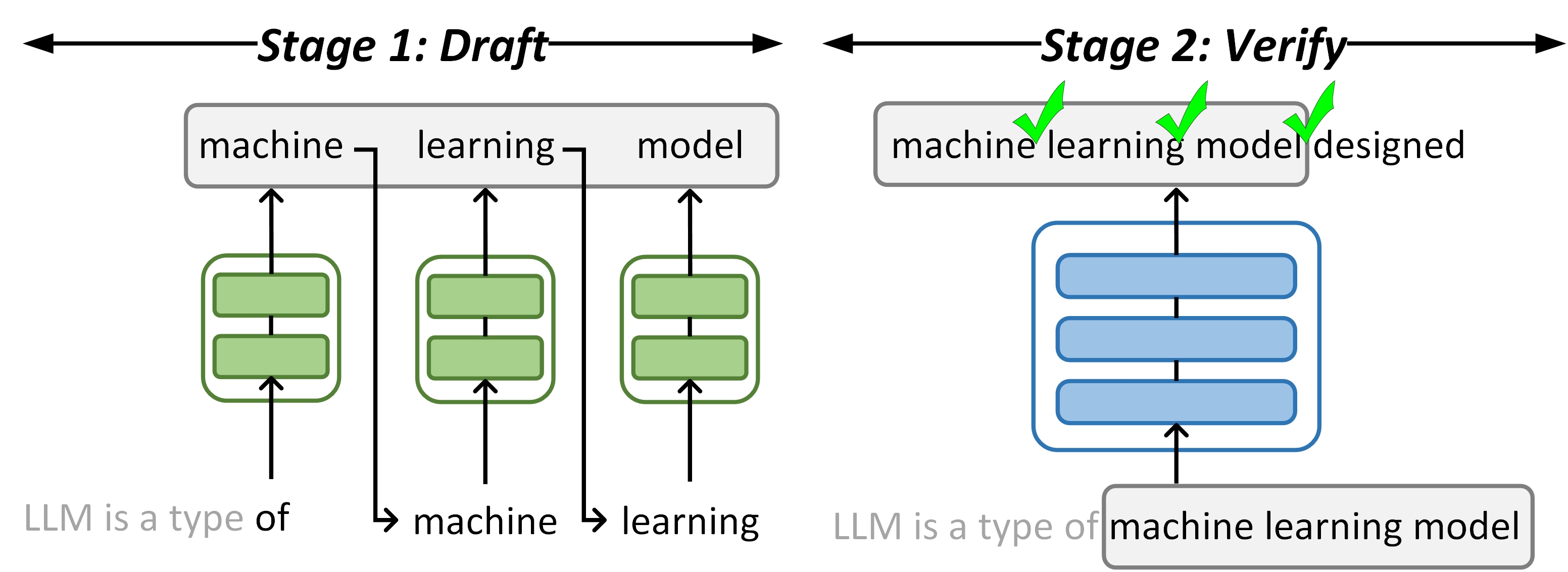} 
    \caption{Illustration of the speculative decoding process: a draft model generates multiple draft tokens, and a target model verifies the draft tokens in parallel.}
    \label{fig_speculativedecoding}
\end{figure}

To accelerate LLM decoding in a lossless manner, speculative decoding has been introduced as an effective strategy \cite{DBLP:conf/asplos/MiaoOZCWZWZYSSC24,DBLP:conf/iclr/Xia00DL25, DBLP:conf/acl/Zhang00S0CM24,DBLP:conf/icml/LiW0024, DBLP:conf/icml/CaiLGPLCD24}. The methodology leverages a fast but less accurate draft model to autoregressively generate a sequence of future tokens. These draft tokens are subsequently verified in a single, parallel forward pass by the original target model, which allows for multiple tokens to be decoded per verification step, thereby speeding up inference without altering the final output (Fig. \ref{fig_speculativedecoding}).
While speculative decoding has garnered significant attention, its practical application is hindered by two persistent challenges: \textbf{(1) the difficulty of creating a high-quality draft model} and \textbf{(2) the suboptimal hardware utilization due to poor co-optimization of the draft and target models.}
Current approaches to speculative decoding typically utilize a trained small auxiliary model or extra heads as the draft \cite{DBLP:conf/asplos/MiaoOZCWZWZYSSC24,DBLP:conf/icml/LiW0024, DBLP:conf/icml/CaiLGPLCD24}, which incurs considerable training burdens and requires additional memory for storage. Self-speculative decoding variants address the training cost by deriving a sparse draft model from the target model itself. However, the quality of this draft is often suboptimal \cite{DBLP:conf/acl/Zhang00S0CM24,DBLP:conf/iclr/Xia00DL25}. Critically, current research predominantly concentrates on GPU-centric software optimizations, failing to explore the co-optimization of the draft-verification pipeline on specialized hardware.

In this paper, we introduce SPEQ, an algorithm-hardware co-design for lossless LLM decoding acceleration, as depicted in Fig. \ref{fig_overview}(d). SPEQ utilizes a self-speculative decoding framework where the draft model is a quantized version of the target model, created via \textbf{a novel bit-sharing quantization algorithm}. This method converts standard FP16/BF16 models into a Bit-Sharing Floating Point (BSFP) format, from which a 4-bit draft model is extracted. This approach eliminates both training requirements and additional memory overhead.
Furthermore, we co-design a \textbf{reconfigurable PE array} optimized to handle the forward passes for both the 4-bit quantized drafting and full-precision verification, ensuring high hardware utilization for the two passes. The key advantages of SPEQ are improved draft token quality, zero memory overhead, no additional training overhead, and hardware co-optimization.
Our evaluations on 15 LLMs and challenging generation tasks show that SPEQ achieves an average draft token accept rate of 0.976. SPEQ accelerator achieves \(2.07\times\), \(1.53\times\), and \(1.45\times\) speedup compared over FP16 baseline, Olive \cite{DBLP:conf/isca/0003THL00LG023}, and Tender \cite{DBLP:conf/isca/LeeLS24}, respectively.

Our contributions are summarized as follows:
\begin{itemize}
    \item We analyze the bit-level representation of the FP16 weights in state-of-the-art LLMs and identify the bit-level underutilization of the exponent part.
    \item We propose bit-sharing quantization, which, to the best of our knowledge, is the first LLM quantization algorithm that enables bit-level sharing between the quantized model and the full FP16 model.
    \item We propose the SPEQ accelerator, a hardware design optimized for efficient LLM speculative decoding. It enables the efficient processing of both the quantized draft forward pass and the full verification forward pass.
    \item We demonstrate the effectiveness of SPEQ across 15 typical LLMs and generative tasks, achieving high acceptance rates and substantial speedups.
\end{itemize}

\section{Background}

\subsection{LLM Quantization}

As depicted in Fig. \ref{fig_overview}(a), for models such as Llama3.1-8B, Llama2-13B, and Llama2-7B \cite{DBLP:journals/corr/abs-2307-09288}, under a representative scenario with a prefill length of 1024 tokens and a decode length of 1024 tokens, memory access for model weights constitutes 98.8\% of the total memory operations during decoding. This underscores the importance of optimizing access of weight memory in the decode stage to mitigate overall latency.

Quantization \cite{DBLP:conf/mlsys/0002TTYCWXDG024, DBLP:journals/corr/abs-2210-17323, DBLP:conf/isca/0003THL00LG023, DBLP:conf/isca/LeeLS24,DBLP:conf/icml/XiaoLSWDH23} is a technique employed to reduce the memory footprint of model weights. It utilizes lower bit-width representations for weights, such as INT8 or INT4, to replace the standard FP16 format. Despite its effectiveness in model compression, several challenges limit the widespread real-world application of quantization.
Firstly, while state-of-the-art 4-bit quantization methods achieve nearly lossless performance on discriminative tasks, \textbf{their efficacy on complex generative tasks remains unsatisfactory}. For instance, \cite{DBLP:journals/corr/abs-2501-03035} demonstrated that applying 4-bit AWQ to Llama3.1-8B-instruct \cite{DBLP:journals/corr/abs-2407-21783} results in a 5.4\% accuracy degradation on the MATH dataset \cite{DBLP:conf/nips/HendrycksBKABTS21}. 
Furthermore, the altered output distribution of a quantized model can introduce \textbf{safety and alignment issues}. This discrepancy necessitates additional fine-tuning to ensure the model does not produce harmful or unintended content, which is a critical requirement for deployment in real-world applications \cite{DBLP:conf/nips/EgashiraVSHV24}.

Although quantization is a powerful method for compressing LLMs, these limitations hinder its broader adoption. Consequently, there is growing interest in exploring lossless methods for accelerating LLM decoding, such as speculative decoding, which aims to reduce latency without compromising model accuracy or safety.

\subsection{Speculative Decoding}

Speculative decoding \cite{DBLP:conf/asplos/MiaoOZCWZWZYSSC24,DBLP:conf/iclr/Xia00DL25, DBLP:conf/acl/Zhang00S0CM24,DBLP:conf/icml/LiW0024, DBLP:conf/icml/CaiLGPLCD24} has emerged as a prominent lossless acceleration technique. This method employs a smaller, faster ``draft" model to generate a sequence of candidate tokens, which are then validated in a single, parallel forward pass by the larger, original ``target" model. The target model determines which tokens from the drafted sequence are correct and can be accepted, while discarding the rest, as illustrated in Fig. \ref{fig_speculativedecoding}. 

Following the analysis in \cite{DBLP:conf/icml/LeviathanKM23}, we conduct a theoretical study on the speedup achieved through speculative decoding. Suppose the draft length of the draft model is \(L\), the time for drafting a single token is \(T_d\), the accept rate is \(r\), the time for verification is \(T_v\). The time for the target model to auto-regressively generate a single token is \(T_{ar}\). The average accept length \(L_a\) can be estimated as:
\begin{equation}
L_a = \frac{1-r^{L+1}}{1-r}
\end{equation}
The speedup over standard autoregressive decoding is given by the following expression. 
We refer the readers to \cite{DBLP:conf/icml/LeviathanKM23} for more details about the speculative decoding process.
\begin{equation}
speedup = \frac{L_a \times T_{ar}}{L\times T_d +T_v} \approx \frac{L_a}{L\times \frac{T_d}{T_{ar}} + 1}
\end{equation}

Recent speculative decoding algorithms for large language models (LLMs) can be broadly classified into two main categories, as shown in Fig. \ref{fig_overview}(b). The first category employs either a smaller LLM or additional output heads as a draft model. For example, SpecInfer \cite{DBLP:conf/asplos/MiaoOZCWZWZYSSC24} trains compact LLMs through knowledge distillation, while Medusa \cite{DBLP:conf/icml/CaiLGPLCD24} trains multiple ``Medusa heads" that generate candidate future tokens based on the model’s final hidden states. However, the Medusa architecture presents significant drawbacks: these heads necessitate \textbf{substantial supplementary training} and introduce an \textbf{approximate 11\% memory overhead} for a Vicuna-7B model instance. Such overheads limit the practicality of these methods in real-world scenarios where computational and storage resources are constrained.

The second category is self-speculative decoding. This approach eliminates the need for additional training and memory resources while improving the quality of the draft tokens. Techniques such as Draft\&Verify \cite{DBLP:conf/acl/Zhang00S0CM24} utilize offline Bayesian optimization to search for a fixed set of sparse layer combinations. Swift \cite{DBLP:conf/iclr/Xia00DL25} employs dynamic layer pruning to construct the draft model. However, \textbf{the pruned model often suffers from severe quality degradation}. For example, Fig. \ref{fig_overview}(b) shows that pruning half of the layers in Llama3.1-8B results in a perplexity increase exceeding \(10^3\).

Despite the progress made in this area, current speculative decoding methods still exhibit significant limitations, including \textbf{the need for additional training and extra memory storage}, as well as the \textbf{low quality of the draft tokens}.
To the best of our knowledge, no prior work has investigated the use of quantized models as draft models. This is primarily because the bit-level representation of commonly quantized models differs from that of the original FP16 models, which requires additional storage for the draft model.

\section{SPEQ Algorithm}

\subsection{Bit-level Underutilization in LLM Weights} \label{sec3.1}


The weights of LLMs are typically stored in floating‑point format, which represents a value as
\begin{equation}
(-1)^{sign} \times 2^{exponent - bias} \times (1 + mantissa)
\end{equation}
In the widely used FP16 format, the exponent field consists of 5 bits, allowing for exponent values in the range \([0,31]\). We analyze the exponent distribution of the weights in several state-of-the-art LLMs, including Llama2-13b \cite{DBLP:journals/corr/abs-2307-09288}, Llama3.1-8b \cite{DBLP:journals/corr/abs-2407-21783}, Qwen2.5-7b \cite{DBLP:journals/corr/abs-2412-15115}, and Vicuna-7b \cite{DBLP:conf/nips/ZhengC00WZL0LXZ23}. As shown in Fig.~\ref{fig_overview}(c), the exponent values across nearly all linear weight tensors are confined to the range 
[0,15]. Virtually no weights exhibit exponents in the range [16,31], suggesting that one exponent bit remains effectively unused.
This underutilization arises from the training practices adopted in modern LLMs. Techniques such as weight decay and normalization are commonly employed to stabilize training and constrain the parameter values, which results in a limited dynamic range of the values and bit-level inefficiency. 


While the above pattern is broadly consistent, certain rare cases exhibit exceptionally large values within specific tensors. For instance, the LLaMA2-13B model exhibits an outlier: a single value of 2.4062, located in the down\_proj tensor of the 0th layer. This value is the sole instance across the entire model to exceed the typical range.
To handle such rare outliers, we introduce a per‑tensor scaling strategy that rescales the weights as needed. The procedure is described in Algorithm \ref{alg3}. This tensor-wise post-scaling operation incurs no performance degradation and negligible computational overhead.

\begin{algorithm}[b]
	\caption{Handling Rare Outliers} 
	\label{alg3} 
	\begin{algorithmic}

	\renewcommand{\algorithmicrequire}{\textbf{Input:}}
	\renewcommand{\algorithmicensure}{\textbf{Output:}}
        \REQUIRE  FP16 weight W
        \STATE $scale = 1.0$
		\IF{$max(abs(W)) >2.0$}
        \STATE $scale = \frac{1.999}{W_{max}}$
        \STATE $W = W*scale$
        \ENDIF

        \STATE $W_q,W_r = BSFP quantize(W)$
        \ENSURE  $W_q,W_r,scale$

	\end{algorithmic} 
\end{algorithm}

\subsection{Bit-sharing Quantization} \label{sec3.2}

In this section, we introduce the BSFP data type, which contains a 4-bit quantized value that shares the bit representation with the full FP16 model. 

\begin{figure}[t]
    \centering
    \includegraphics[width=0.95\columnwidth,keepaspectratio]{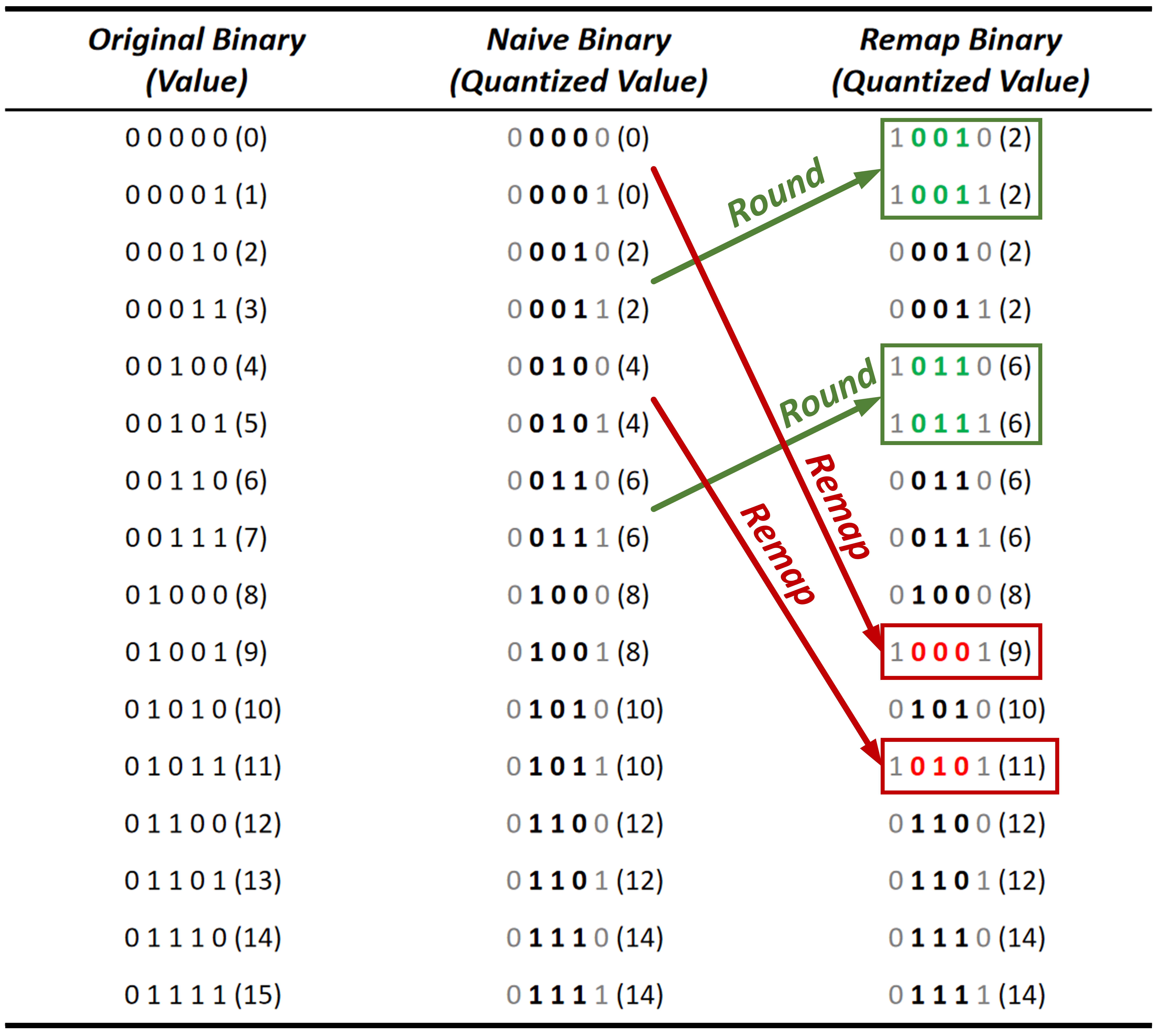} 
    \caption{Illustration of remapping: the first row "Original" means the binary encoding and real value of FP16 exponent. "Naive" means using naive E3M0 quantization, the two neighbor values are rounded to the same quantized value. The bits in bold mean the binary representation of the quantized value. "Remap" means remapping the binary encoding to create unique representations for 9 and 11, while 0-3 and 4-7 are rounded to the same quantized value. The wasted bit is set to 1 if the bit representation of the remapped value is different from the original value.}
    \label{fig_remap}
\end{figure}

To create the bit-sharing property, the 4-bit quantized model needs to be in FP4 datatype. There are various FP4 representations, including E1M2, E2M1, and E3M0. We perform fine-grained group quantization with a group size of 128. 
Using E3M0 as an example, as shown in the middle part of Fig. \ref{fig_remap}, for each group, we first extract the middle 3 bits of the exponent part of the FP16 values \(w_i\) to obtain the quantized values \(Q(w_i)\) (the values in bold), then we compute a scale \(s\) to minimize the mean square error between the quantized values and the original values according to Equation \ref{scale}. During inference, we multiply the scale \(s\) with the final output of the quantization group.
\begin{equation}
\label{scale}
s=\frac{\sum_{i=0}^{127}w_i \times Q(w_i)}{\sum_{i=0}^{127}Q(w_i)^2}
\end{equation}
We evaluate all possible types of FP4 representations on wikitext2 \cite{DBLP:conf/iclr/MerityX0S17} with a sequence length of 2048, and from Table \ref{table2}, it is observed that using the E3M0 format for representing the quantized model results in the lowest perplexity for all three models, since the exponent part has a more significant impact than the mantissa part.

\begin{table}[b]
\centering
\caption{Perplexity results of different quantization methods for the draft model with bit-sharing.}
\label{table2}
\begin{tabular}{c c c c}
\toprule[1pt]
\textbf{Method} & \makecell{Llama3.1-8b} & \makecell{Llama2-7b} & \makecell{Vicuna-7b} \\
\midrule
FP16 & 6.2 & 5.5 & 6.8 \\
\midrule
E1M2 & 3E+5 & 2E+4 & 3E+4 \\
E2M1 & 7E+4 & 7E+3 & 1E+4  \\
E3M0 & 251.8 & 153.9 & 103.0 \\
\midrule
Naive & 251.8 & 153.9 & 103.0 \\
\textbf{+Remap} & \textbf{10.5} & \textbf{7.0} & \textbf{8.9}  \\

\bottomrule[1pt]
\end{tabular}
\end{table}

\begin{figure*}[htbp]
    \centering
    \includegraphics[width=0.95\textwidth,keepaspectratio]{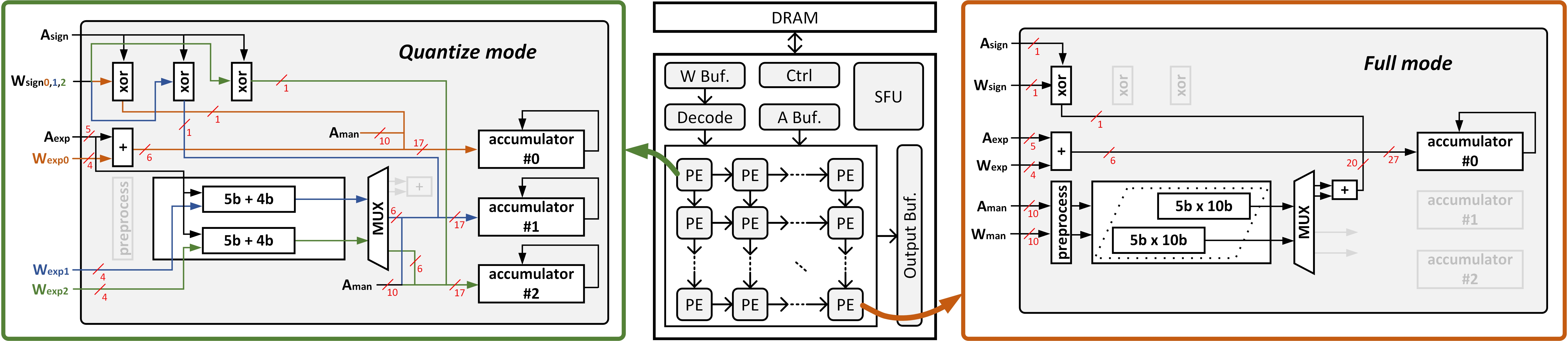} 
    \caption{Overview of SPEQ hardware accelerator. The reconfigurable PE is capable of processing both the quantized GEMM and the full GEMM with high hardware efficiency.}
    \label{fig_hwoverview}
\end{figure*}

The key feature of BSFP is to utilize the wasted bit while better preserving the important values in the exponent. As shown in Table \ref{table2}, applying E3M0-based shared quantization without modification results in a significant increase in perplexity for the three models. This occurs because the nearest two exponent values are rounded to the same value, introducing substantial errors in the important large values.

To address this issue, we propose a novel remapping method, illustrated in Fig. \ref{fig_remap}.
As shown in Fig. \ref{fig_overview}(c), values in the range of 8-11 are more critical than values in the range of 0-7 due to their larger magnitudes and quantities, which are essential for maintaining numerical precision. 
Therefore, instead of rounding 8 and 9 to 8, and 10 and 11 to 10, we preserve separate representations for 9 and 11 by assigning unique encodings, specifically 3'b000 and 3'b010. The original values represented by 3'b000 and 3'b010 are then rounded to 3'b001 and 3'b011, respectively. 
This remapping improves the precision of the quantized values but compromises the bit-sharing characteristic. To recover from this, we use the unused bit for correction. If a value’s bit representation changes due to remapping, we set the unused bit to 1, signaling that this value needs to be looked up. The values in bold in Fig. \ref{fig_remap} right represent the binary representation of the quantized values, which are stored in the quantized weight with the sign bit, denoted as \(W_q\). The remaining values, including 2 exponent bits and 10 mantissa bits are concatenated and denoted as \(W_r\). Using our remapping method gains a significant performance boost for LLMs, as illustrated in Table \ref{table2}, resulting in a significantly smaller perplexity increase than using naive E3M0 quantization.



\subsection{Optimization for Speculative Decoding}
We incorporate an early exit mechanism in the drafting process. If the predicted probability of the draft model is low, it suggests that the token is unlikely to pass verification, prompting us to stop drafting and avoid wasting resources on tokens that are likely to be rejected. We set a parameter \(\gamma\) for early stopping. 
If the sampled maximum draft logit  $p_{draft}(x_{i+1}|x_1, ..., x_{i})<\gamma$, the drafting process is halted, and verification begins. Otherwise, drafting continues until the maximum draft length \(L\) is reached.
Note that SPEQ incurs no extra KV cache overhead, as the quantized draft model and the full model share the same KV cache.

\section{SPEQ Hardware Accelerator}
\subsection{Overview} \label{sec4.1}

Fig. \ref{fig_hwoverview} presents the overall architecture of SPEQ accelerator. It is composed of DRAM, on-chip buffers, a control unit, a special function unit (SFU), and a reconfigurable PE array. The control unit manages data movement between DRAM and the on-chip buffer, and controls the processing of the PE array. The on-chip buffers contain a 512 KB W buffer, a 512 KB A buffer, and a 512 KB output buffer. The SFU processes the activation functions. The reconfigurable PE array contains \(32 \times 32\) PE units, which are capable of handling quantized forward and full forward. 

The SPEQ hardware accelerator is primarily designed for FP16 processing, meaning that during the full forward pass, both the activations and weights are represented in FP16. \textbf{This accelerator is also capable of supporting BF16 models}. For models represented in the BF16 format, we first round the exponent values that are smaller than 112 up to 112. Subsequently, a similar remapping process is applied to the exponent component. In this manner, the exponent part of the BF16 representation is successfully reduced to 5 bits with bit-sharing property. Furthermore, we pad the mantissa component with three zeros. This results in weights being represented in the same format as FP16 (S1E5M10), allowing us to leverage the quantized forward and full forward passes within the PE array.

\subsection{Decoder for BSFP} \label{sec4.4}
In Section \ref{sec3.2}, we introduced the remapping scheme in BSFP to enhance the performance of the draft model. This remapping results in a different bit representation compared to standard FP4 and FP16, necessitating the use of special decoders to encode the values for subsequent computation.

\begin{figure}[t]
    \centering
    \includegraphics[width=0.95\columnwidth,keepaspectratio]{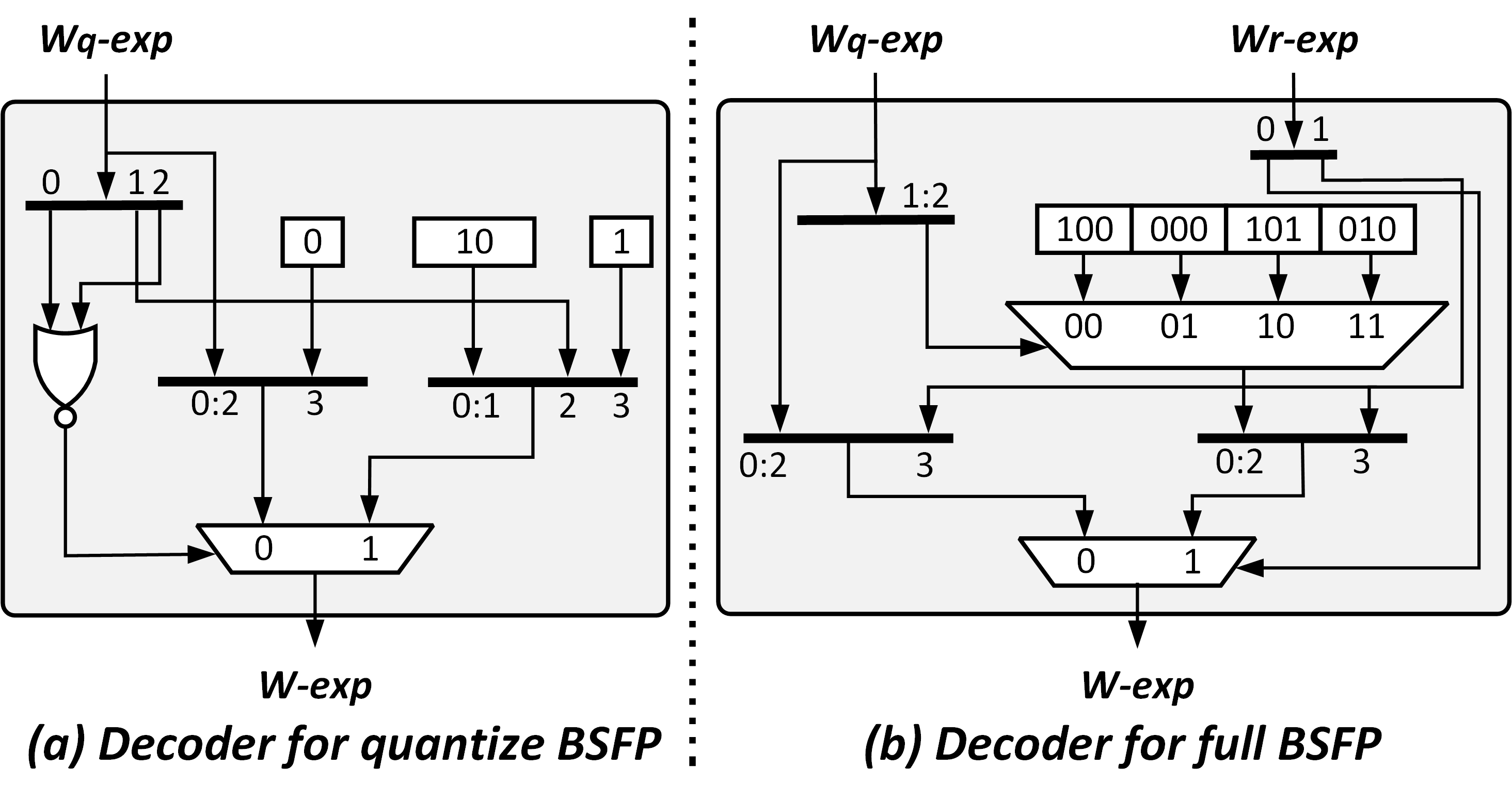} 
    \caption{Decoder design for BSFP.}
    \label{fig_decoder}
\end{figure}

The decoder design for the exponent of the quantized value \(W_q\) is shown in Fig. \ref{fig_decoder}(a). The 3 bit  \(W_{q-exp}\) is input into the decoder unit. As illustrated in Fig. \ref{fig_remap}, the values that need to be looked up are 9 and 11, represented in binary as 3'b000 and 3'b010, respectively. 
We first extract the 0th and 2nd bits and pass them through a NOR gate. If the output is 0, indicating that no decoding is required, a zero is appended to the end of the value to produce the output. If the output is 1, indicating that the value corresponds to 9 or 11, the final binary representations are 4'b1001 or 4'b1011. We set the 0th, 1st, and 3rd bits to 1, 0, and 1, respectively, and use the 1st bit of \(W_{q-exp}\) as the 2nd bit of the output. In this way, we obtain the final output of 4 bits. 

The decoder design for the exponent of the FP16 value is shown in Fig. \ref{fig_decoder}(b). The 3-bit \(W_{q-exp}\) and 2-bit \(W_{r-exp}\) from \(W_r\) are used as inputs. If the 0th bit of \(W_{r-exp}\) is 0, meaning that the weight remains unchanged, the two parts are concatenated to obtain the final output. If the 0th bit of \(W_{r-exp}\) is 1, we use the 1st and 2nd bits of \(W_{q-exp}\) for the MUX unit, the MUX takes in 2 bits and outputs 3 bits as shown in Fig. \ref{fig_decoder}. 
The output value from the MUX is then concatenated with the 0th bit of \(W_{r-exp}\) to obtain the final result. 

\subsection{Reconfigurable PE Array} \label{sec4.5}
The PE array in the SPEQ accelerator adopts a reconfigurable architecture, consisting of 8 tiles with 128 PEs each. During drafting, each PE operates in quantized mode, processing three 5-bit weights and one FP16 activation to generate three partial sums. During verification, it processes one FP16 weight and one FP16 activation to produce a single partial sum.

Each PE includes four main components: (1) a sign processing unit that handles sign bits via XOR operations; (2) an adder unit that combines activation and weight exponents; (3) a multiplication unit employing a Wallace tree for mantissa operations; and (4) an accumulation unit that performs FP32 accumulation on the weighted products.

\begin{figure}[t]
    \centering
    \includegraphics[width=0.95\columnwidth,keepaspectratio]{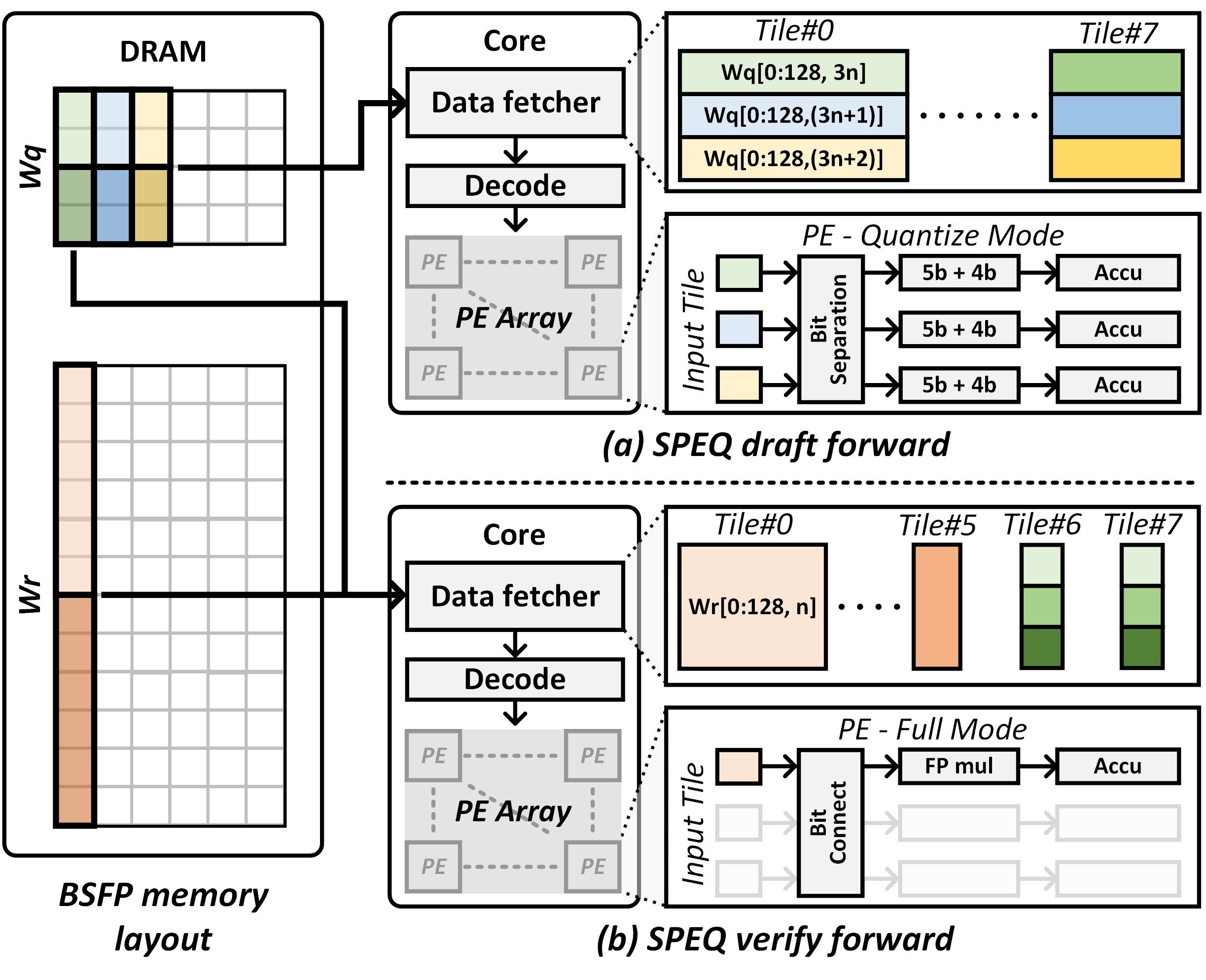} 
    \caption{Workflow of the quantize mode and full mode of SPEQ hardware.}
    \label{fig_workflow}
\end{figure}

The workflow for quantized inference and full inference is shown in Fig. \ref{fig_workflow}. 
In full mode, the processing of the PE array is identical to that of a standard FP16 PE array. However, the actual bit-width per element is 15 bits, as the unused exponent bit is ignored after decoding. This results in a total input bit-width of 16 (activation) + 15 (weight) = 31 bits.
Within the multiplication unit, the 10-bit mantissa of the weight is divided into two 5-bit segments: the upper and lower halves. Each segment is multiplied with the 10-bit mantissa of the activation value using two parallel Wallace trees. The outputs of these two multiplications are then summed to produce a 20-bit product. Simultaneously, the exponents of the weight and activation are added using a 5-bit adder tree. Finally, the sign, exponent, and mantissa are concatenated and passed to the accumulation unit, where FP32 accumulation is performed.

In quantize mode, each PE tile receives \(3\times128\) quantized weight elements as shown in Fig. \ref{fig_workflow}. The weights of three different weight groups corresponding to the same activation are processed by a single PE. Each weight group undergoes its own MAC operation to obtain the final output. Afterward, the results are accumulated in a single PE tile. 
The PE in quantize mode receives an activation and three quantized weights. The total input bit-width is 16 (activation) + 3 \(\times\) 5 (weight) = 31 bits, the same as in full mode. 
Each of the three quantized weights consists of 1 bit for the sign and 4 bits for the exponent. For the first weight, the exponent is added to the activation exponent in the exponent adder of the PE. For the second and third weights, we utilize the components in the mantissa multiplier to perform the addition. The two adders in the two Wallace trees function as separate adders during this process, while the remaining components in the multiplier are masked out. The resulting added exponents of the activation and the three weights are concatenated with the mantissa part of the activation and its sign. These are then fed into three accumulation units to perform accumulation separately.

\section{Evaluation}

\subsection{Experiment Setup}
\textbf{LLM Benchmarks.} For the evaluation, we utilize SOTA LLMs including Llama3.1-8b \cite{DBLP:journals/corr/abs-2407-21783}, Vicuna-7b \cite{DBLP:conf/nips/ZhengC00WZL0LXZ23}, Llama2-7b \cite{DBLP:journals/corr/abs-2307-09288}, Llama3.2-3b \cite{DBLP:journals/corr/abs-2407-21783}, and Llama2-13b \cite{DBLP:journals/corr/abs-2307-09288}. The evaluation is conducted across 3 generative tasks: MT-bench, GSM8K, and Humaneval. 
\begin{itemize}
    \item \textbf{GSM8K \cite{DBLP:journals/corr/abs-2110-14168}} includes 8,000 high-quality grade-school mathematics problems designed to evaluate the mathematical reasoning capabilities of LLMs. We use 8-shot prompt for GSM8K evaluation.
    \item \textbf{Humaneval \cite{DBLP:journals/corr/abs-2107-03374}} is a benchmark focused on assessing the code generation capabilities of LLMs. 
    \item \textbf{MT-Bench \cite{DBLP:conf/nips/ZhengC00WZL0LXZ23}} is a benchmark designed to evaluate the multi-turn conversational abilities of LLMs.
\end{itemize}

\noindent \textbf{Hardware Implementation.} We complete the synthesis, placement, and routing of SPEQ accelerator based on 28nm CMOS technology. We obtain the working power based on VCS and Verdi simulation.

\noindent \textbf{Hardware Accelerator Baselines.} We compare the speedup and energy efficiency between SPEQ and existing quantization-based hardware accelerators, including Olive\cite{DBLP:conf/isca/0003THL00LG023} and Tender\cite{DBLP:conf/isca/LeeLS24}. 
Note that Olive and Tender result in substantial increases in perplexity and accuracy degradation. For the Llama2-7b model, \textbf{4-bit Olive leads to a perplexity increase of 38.7} (\(5.5 \rightarrow 44.2\)), while \textbf{4-bit Tender causes a perplexity increase of 31.0} (\(5.5 \rightarrow 36.5\)).
The \textbf{severe performance degradation} associated with these methods limits their applicability, whereas our approach is \textbf{lossless}. We also include the results for 8-bit Olive and Tender with better accuracy performance.

\begin{table}[t]
\centering
\caption{Average draft length and accept rate of 5 LLMs on 3 generative tasks.}
\resizebox{\columnwidth}{!}{
\begin{tabular}{lccccccc}
\toprule[1pt]
\multirow{2}{*}{\textbf{Model}} & \multicolumn{2}{c}{\textbf{Humaneval}} & \multicolumn{2}{c}{\textbf{MT-bench}} & \multicolumn{2}{c} {\textbf{GSM8K}} & \multicolumn{1}{c}{\textbf{Mean}} \\

\cmidrule(lr){2-3} \cmidrule(lr){4-5} \cmidrule(lr){6-7}
 & \(L\) & \(r\) & \(L\) & \(r\) & \(L\) & \(r\)  & \(r\) \\
\midrule

Vicuna-7b  & 8.02  & 0.968  & 8.40 & 0.964  & 7.48 & 0.977  & 0.970   \\

Llama2-7b  & 6.05 & 0.981  & 4.47 & 0.986  & 6.38 & 0.987  & 0.985  \\

Llama3.1-8b & 5.10 & 0.975  & 5.69 & 0.979 & 5.31 & 0.967 & 0.974 \\

Llama3.2-3b & 5.61 & 0.953 & 6.05 & 0.978 & 4.83 & 0.964 & 0.965\\

Llama2-13b & 5.80 & 0.986 & 6.61 & 0.992 & 6.57 & 0.991 & 0.990 \\

\bottomrule[1pt]
\end{tabular}
\label{table3}
}
\end{table}

\subsection{LLM Benchmarks}
We report the average draft length \(L\), accept rate \(r\), and speedup on the 3 benchmarks. Note that our method is a lossless acceleration method, the accuracy is the same as the full model. For all tasks, we set the generation length to 256.
We use a maximum draft length of 16 and an early stopping threshold \(\gamma\) of 0.6. Further evaluation of hyperparameter settings is in Section \ref{sec5.4}.

Table \ref{table3} presents the average draft length and accept rate of SPEQ. SPEQ achieves an average accept rate of 0.977 for the 5 LLMs, indicating the high quality of the draft model and the high alignment between the quantized draft model and the target model. 

\begin{table}[t]
\centering
\caption{Speedup of 5 LLMs on 3 generative tasks.}
\begin{tabular}{lcccc}
\toprule[1pt]
{\textbf{Model}} & {\textbf{Humaneval}} & {\textbf{MT-bench}} & {\textbf{GSM8K}} & {\textbf{Mean}} \\

\midrule
Vicuna-7b  &  2.05\(\times\) &  2.03\(\times\) &  2.12\(\times\) &  2.07\(\times\)  \\
Llama2-7b  &  2.11\(\times\)  & 2.04\(\times\)  & 2.16\(\times\) &  2.10\(\times\) \\
Llama3.1-8b & 2.01\(\times\) &  2.08\(\times\)  & 2.00\(\times\) & 2.03\(\times\) \\

Llama3.2-3b & 1.93\(\times\) & 2.09\(\times\) & 1.96\(\times\) & 2.00\(\times\) \\

Llama2-13b & 2.13\(\times\) & 2.21\(\times\) & 2.19\(\times\) & 2.18\(\times\) \\
\bottomrule[1pt]
\end{tabular}
\label{table4}
\end{table}

Table \ref{table4} shows the speedup results of SPEQ compared with FP16 baseline. SPEQ achieves an average speedup of 2.08\(\times\) compared with naive FP16 autoregressive decoding for the 5 LLMs.

\subsection{Hardware Accelerator Performance}

\textbf{Area and Energy Evaluation.} Table \ref{table5} shows the power and area breakdown of SPEQ. It has a total area of 6.3 mm\(^2\). The PE array takes up \(39.4\%\) area, the decoder only occupies \(3.5\%\) of the total area, which is a small overhead.
In the quantize mode, the PE array takes up \(36.5\%\) energy. The decoder unit only occupies \(3.2\%\) power. In the full mode, the PE array takes up \(40.0\%\) energy. The decoder unit only occupies \(3.1\%\) power.
When working at 500 MHz, the power of quantize mode is 508 mW, and the power of full mode is 559 mW. The similar power consumption of the quantize mode and full mode indicates high hardware utilization of SPEQ accelerator in both the two modes.

\begin{table}[t]
\centering
\caption{Area and power breakdown of SPEQ accelerator at 500 MHz.}
\label{table5}
\begin{tabular}{lccc}
\toprule[1pt]
\textbf{Module} & \textbf{Area} & \makecell{\textbf{Power}\\(quantize mode)} & \makecell{\textbf{Power}\\(full mode)}\\ 
\midrule
PE & 39.4\% & 36.5\% & 40.0\%\\ 
Decoder & 3.5\% & 3.2\% & 3.1\% \\ 
SRAM & 35.1\% & 32.1\% & 30.2\%\\ 
VPU & 14.8\% & 15.3\% & 14.5\% \\
Others & 7.2\% & 12.9\% & 12.2\%\\ 
\midrule
\textbf{Total} & \textbf{6.3$mm^2$} & \textbf{508mW} & \textbf{559mW}\\ 
\bottomrule[1pt]
\end{tabular}
\end{table}

\begin{figure}[b]
    \centering
    \includegraphics[width=\columnwidth,keepaspectratio]{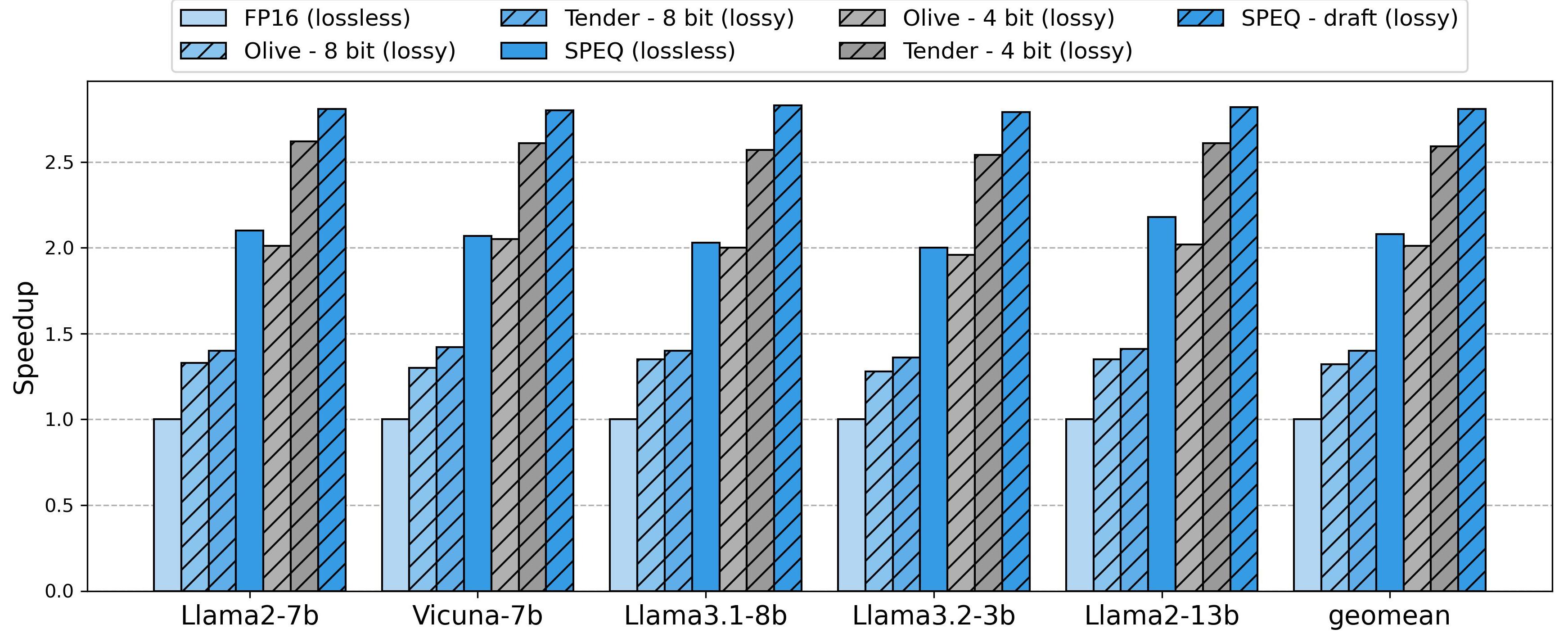} 
    \caption{Speedup of SPEQ compared with FP16 baseline, Olive, and Tender.}
    \label{fig_speedup}
\end{figure}

\begin{figure}[t]
    \centering
    \includegraphics[width=\columnwidth,keepaspectratio]{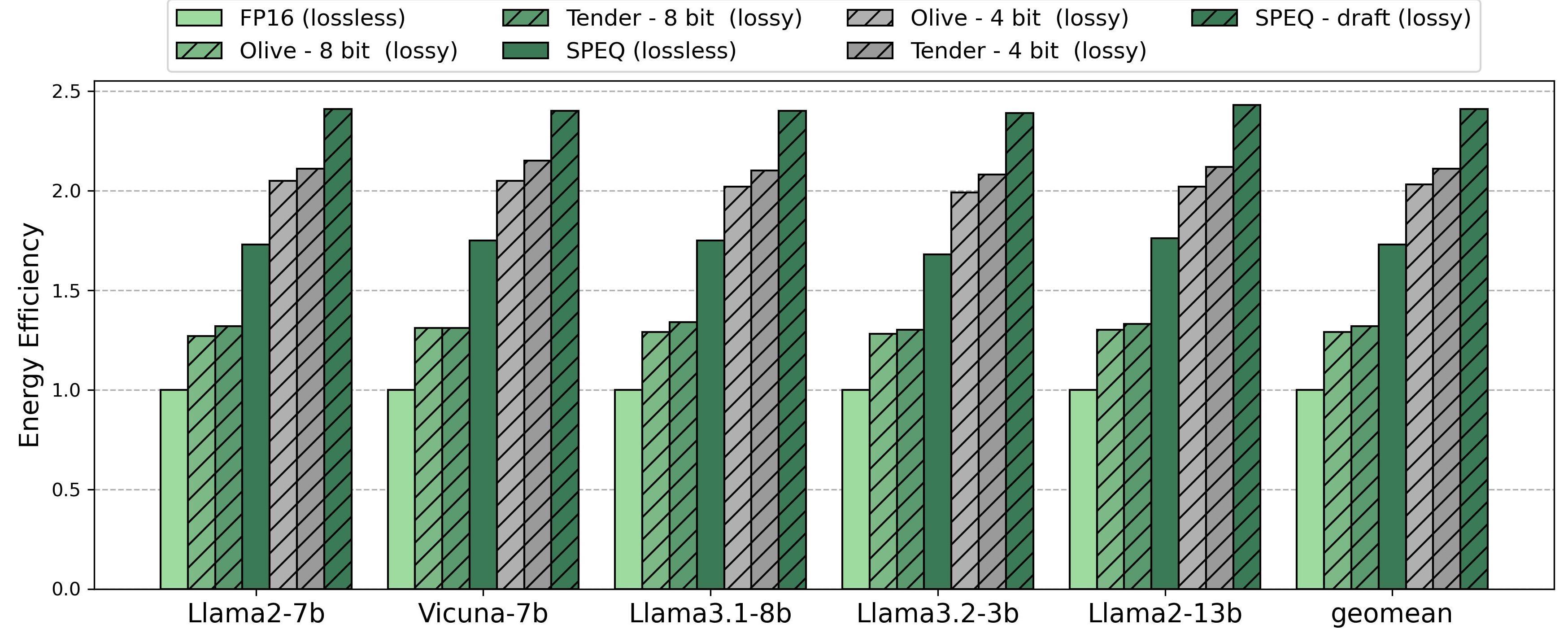} 
    \caption{Energy efficiency of SPEQ compared with FP16 baseline, Olive, and Tender.}
    \label{fig_energy}
\end{figure}

\begin{figure}[t]
    \centering
    \includegraphics[width=\columnwidth,keepaspectratio]{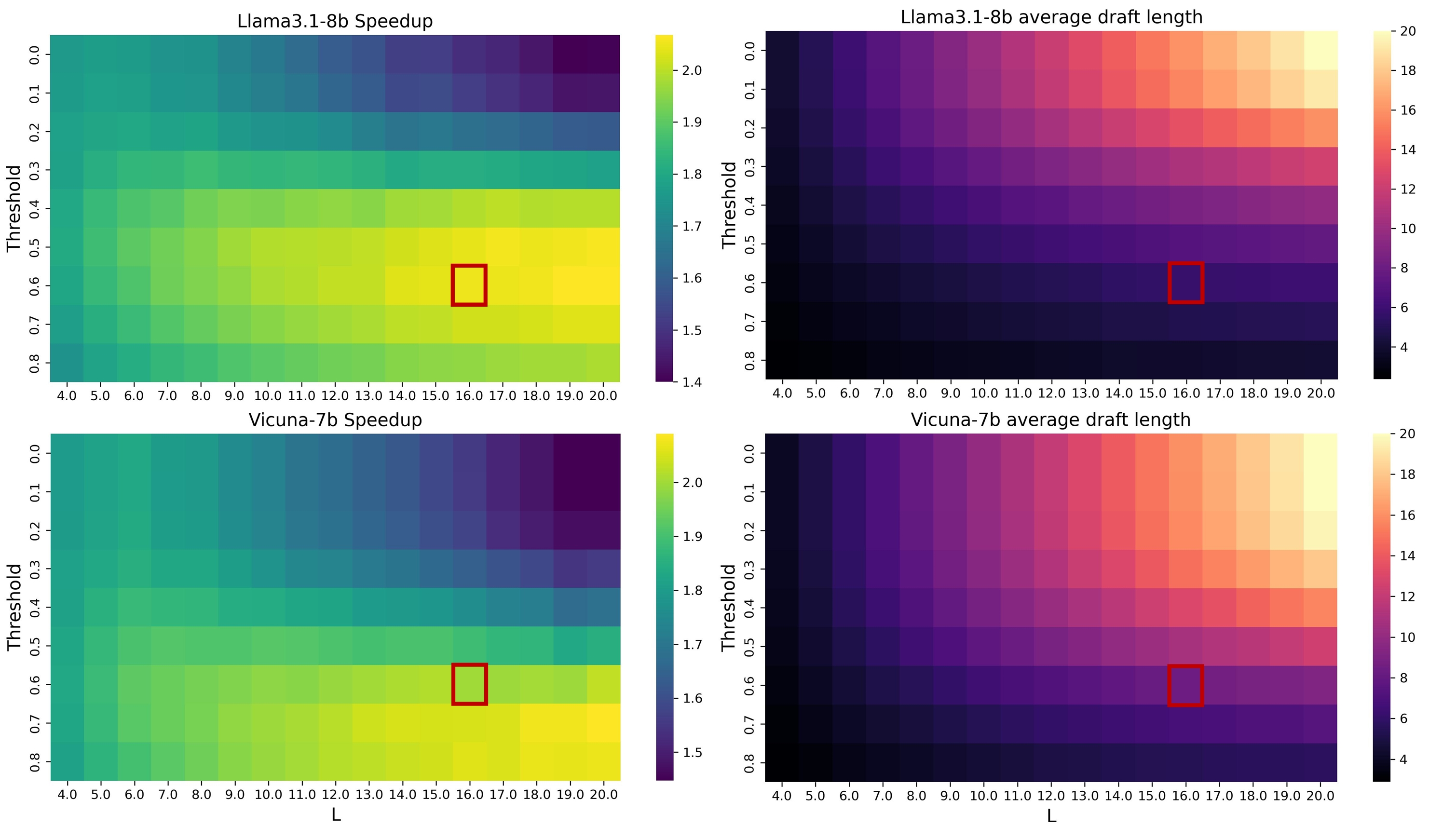} 
    \caption{Ablation study of the hyperparameter settings. The square stands for the default setting of \( L=16, \gamma= 0.6 \).}
    \label{fig_ablation}
\end{figure}

\noindent \textbf{Comparison with Other LLM Quantization Accelerators.} 
The comparison of decoding speedup is shown in Fig. \ref{fig_speedup}. We use the mean speedup of the three tasks as the speedup of SPEQ. Methods exhibiting \textbf{severe performance degradation} (4-bit Olive and 4-bit Tender) are marked in gray.
SPEQ achieves to an average speedup of \(2.07\times\), \(1.53\times\), and \(1.45\times\) compared with the FP16 baseline, 8-bit Olive \cite{DBLP:conf/isca/0003THL00LG023}, and 8-bit Tender \cite{DBLP:conf/isca/LeeLS24}, respectively. Additionally, SPEQ demonstrates a similar speedup when compared with 4-bit Olive.

The superior performance of SPEQ derives from the efficient processing of the quantized forward, where only the exponents are added to perform multiplication between the quantized value and the activation. SPEQ achieves a high accept rate and long draft length, minimizing the latency overhead of the verification process. 

The comparison of energy efficiency is shown in Fig. \ref{fig_energy}. SPEQ achieves \(1.74\times\) energy efficiency compared with the FP16 baseline, \(1.35\times\) and \(1.32\times\) compared with 8-bit Olive and Tender. 


\subsection{Comparison with Other Speculative Decoding Methods}
We further compare SPEQ with other speculative decoding methods, including Swift \cite{DBLP:conf/iclr/Xia00DL25} Medusa \cite{DBLP:conf/icml/CaiLGPLCD24} for Vicuna-7b on MT-bench. 
As shown in Fig. \ref{fig_overview}(a), SPEQ achieves a speedup of \(2.03\times\), \textbf{surpassing Swift by \(1.52\times\) and Medusa by \(1.05\times\) without additional training and extra memory overhead}.

\subsection{Ablation Study on Hyperparameter Settings} \label{sec5.4}

We conduct an ablation study on the hyperparameter settings of the maximum draft length \(L\) and the early stopping threshold \(\gamma\) for Llama3.1-8b and Vicuna-7b on MT-bench. The value of \(L\) ranges from 4 to 20 with an interval of 1, and \(\gamma\) ranges from 0 to 0.8 with an interval of 0.1. 
The results are shown in Fig. \ref{fig_ablation}. Our default setting of \(L=16, \gamma=0.6\) yields significant speedup. The optimal choice of hyperparameters depends on the specific model and task. Although our default settings may not represent the absolute optimal configuration, the results in Table \ref{table4} demonstrate that they achieve performance comparable to the optimal settings.

\section{Conclusion}
In this paper, we propose SPEQ, an algorithm-hardware co-design solution that accelerates LLM decoding without sacrificing accuracy. SPEQ employs a bit-sharing quantization algorithm with a reconfigurable PE array to support both quantized draft and full model verification, providing efficient computation and high hardware utilization. Experiments on 15 LLMs and generative tasks demonstrate that SPEQ outperforms prior lossy quantization accelerators, offering a promising solution for accelerating LLM inference in real-world applications.

\bibliographystyle{IEEEtranS}
\bibliography{main}

\end{document}